\documentclass[10pt,conference]{IEEEtran}

\usepackage{amsmath}
\usepackage{cite}
\usepackage{graphicx}
\usepackage{amsfonts}
\usepackage{latexsym}
\usepackage{subfigure}
\usepackage{algorithm}
\usepackage{algorithmic}
\usepackage{color}
\usepackage{times}
\usepackage{epsfig, graphics}
\usepackage{bm}
\usepackage{subfigure}
\usepackage{graphicx,wrapfig,lipsum}
\usepackage{amsthm}

\begin{document}

\newtheorem{theorem}{Theorem}
\newtheorem{lemma}{Lemma}
\newtheorem{conjecture}{Conjecture}
\newtheorem{corollary}{Corollary}
\newtheorem{definition}{Definition}
\newtheorem{scheme}{Scheme}
\newcommand{\argmax}{\arg\!\max}
\newcommand{\rev}[1]{{\color{red}#1}} 
\newcommand{\pound}{\operatornamewithlimits{\gtrless}}
\IEEEoverridecommandlockouts

\title{Membership Inference Attack and Defense for Wireless Signal Classifiers with Deep Learning}

\author{\IEEEauthorblockN{Yi Shi, \emph{Senior Member, IEEE} and Yalin E. Sagduyu, \emph{Senior Member, IEEE}} 
\thanks{Yi Shi is with Virginia Tech, Blacksburg, VA, USA; Email: yshi@vt.edu. Yalin E. Sagduyu is with Intelligent Automation, Inc., Rockville, MD, USA; Email: ysagduyu@i-a-i.com.
}
\thanks{This effort is supported by the U.S. Army Research Office under contract W911NF-20-C-0055. The content of the information does not necessarily reflect the position or the policy of the U.S. Government, and no official endorsement should be inferred.} \thanks{A preliminary version of the material in this paper was partially presented at ACM Conference on Security and Privacy in Wireless and Mobile Networks (WiSec) Workshop on Wireless Security and Machine Learning (WiseML) \cite{WiseML2020}.}
}

\maketitle

\begin{abstract}
An over-the-air membership inference attack (MIA) is presented to leak private information from a wireless signal classifier. Machine learning (ML) provides powerful means to classify wireless signals, e.g., for PHY-layer authentication. As an adversarial machine learning attack, the MIA infers whether a signal of interest has been used in the training data of a target classifier. This private information incorporates waveform, channel, and device characteristics, and if leaked, can be exploited by an adversary to identify vulnerabilities of the underlying ML model (e.g., to infiltrate the PHY-layer authentication). One challenge for the over-the-air MIA is that the received signals and consequently the RF fingerprints at the adversary and the intended receiver differ due to the discrepancy in channel conditions. Therefore, the adversary first builds a surrogate classifier by observing the spectrum and then launches the black-box MIA on this classifier. The MIA results show that the adversary can reliably infer signals (and potentially the radio and channel information) used to build the target classifier. Therefore, a proactive defense is developed against the MIA by building a shadow MIA model and fooling the adversary. This defense can successfully reduce the MIA accuracy and prevent information leakage from the wireless signal classifier.
\end{abstract}


\section{Introduction}
Machine learning (ML) has emerged with powerful means to learn from and adapt to wireless network dynamics, and solve complex tasks in wireless communications subject to channel, interference, and traffic effects. In particular, \emph{deep learning} (DL) that has been empowered by recent algorithmic and computational advances can effectively capture high-dimensional representations of spectrum data and support various wireless communications tasks, including but not limited to, spectrum sensing, signal classification, spectrum allocation, and waveform design \cite{bc}. However, the use of ML/DL also raises unique challenges in terms of security for wireless systems \cite{aml, aml2}. With \emph{adversarial machine learning} (AML), various attacks have been developed to launch against the ML/DL engines of wireless systems, including inference (exploratory) attacks \cite{Shi2018, terpek, NetSliceAttack}, evasion (adversarial) attacks \cite{Larsson, Larsson2, Yalin2019, Headley2019, Headley2019-1, Headley2020, Silvija2019, Silvija2019-2, Silvija2019-3, Kim, Kim2, Mao2021, Kim4, Kim5, Larsson3, Kim6, sahay2021deep, Bahramali, Jinho}, poisoning (causative) attacks \cite{Sagduyu1, YiMilcom2018, Luo2019, Luo2020, Luo2021}, Trojan attacks \cite{Davaslioglu19}, spoofing attacks \cite{Shi2019, Shi2021, 5Gbc}, and attacks to facilitate covert communications \cite{Gunduz1, Gunduz2, Kim3}. These AML-based attacks operate with small spectrum footprints and thus are harder to detect compared with conventional wireless attacks such as jamming of data transmissions \cite{conventional1, conventional2}.

  \begin{figure}
   \centering
   \includegraphics[width=0.8\columnwidth]{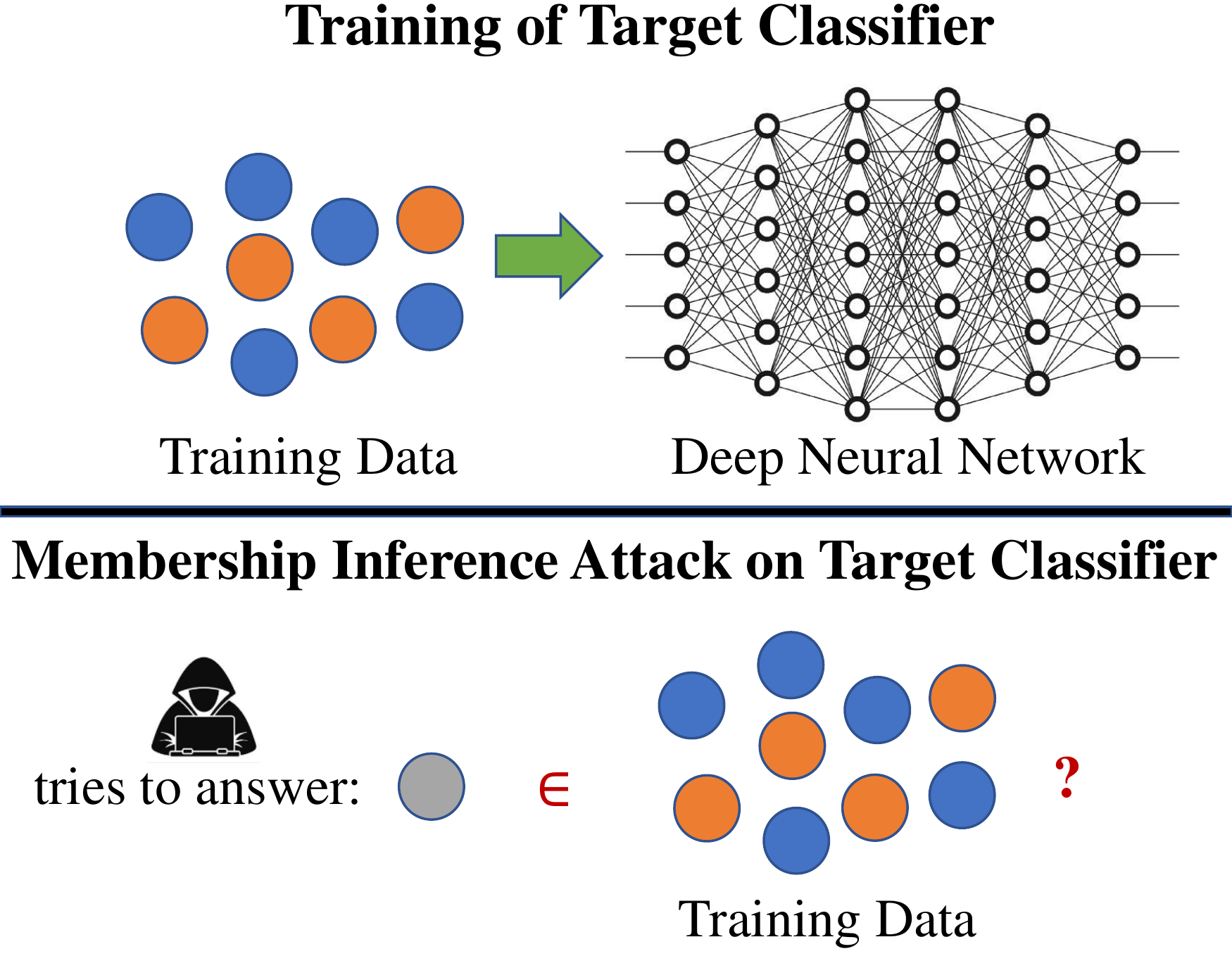}
   \caption{The membership inference attack (MIA).}\label{fig:mia}
 \end{figure}

In conjunction with security threats, an emerging concern on ML-based solutions is \emph{privacy}, namely the potential leakage of information from the ML models to the adversaries. 
One example is the model inversion attack, where the adversary has access to the ML model and some private information, and aims to infer additional private information by observing the inputs and outputs of the ML model \cite{ModelInv}.
Another privacy attack of interest is the \emph{membership inference attack} (MIA) \cite{Shokri17, Nasr18, Yeom18, Salem18, Leino19, Nasr19, Song19, Song19-2, Truex19, Jia20, Song20, Gong21, Dingfan} that has been extensively studied in various data domains including computer vision, healthcare, and commerce. The goal of the MIA to infer if a particular data sample has been used in training data or not (see Fig.~\ref{fig:mia}).  While the MIA has been demonstrated as a major privacy threat for computer vision and other data domains, it has not been applied yet to the wireless domain. In practice, the broadcast and shared nature of wireless medium offers unique opportunities to an adversary to eavesdrop wireless transmissions and launch the MIA \emph{over the air} against a wireless signal classifier to infer about the underlying radio device, waveform, and channel environment characteristics under which the ML/DL model of the target signal classifier is trained. 

In this paper, we present the first application of the MIA in the wireless domain. In particular, we consider a wireless signal classifier based on a \emph{deep neural network} (DNN) as the target ML engine against which the MIA is launched over the air. For the \emph{PHY-layer authentication} of potentially massive number of heterogeneous users (e.g., Internet of Things (IoT) devices), a service provider (e.g., gNodeB in 5G applications such as network slicing) can use such a classifier to classify users as authorized or not based on the RF fingerprints in the received signals (reflecting the inherent characteristics of the user's RF transceiver along with channel effects) \cite{PHY, Dyspan, Monson, Deepwifi} and then admit communication requests of authorized users (e.g., it can be potentially implemented as xApps in the near Real-Time RAN Intelligent Controller (Near-RT RIC) of O-RAN). 
The adversary can sense the spectrum to observe the behavior of a target classifier and then launch the MIA to determine whether a data sample (a wireless signal) of interest is in the training data of the target classifier or not. \emph{This attack reveals whether a wireless signal classifier is trained against a particular waveform, radio device, or channel environment}. This private information leakage can be further exploited by the adversary to launch other attacks. For example, the adversary can spoof signals similar to the ones from authorized users using the same type of radio device and waveform and under a similar spectrum environment. This way, the adversary can bypass the signal classifier trained for PHY-layer authentication, and can gain network access or prevent access of other users by occupying communication resources.

The wireless systems pose \emph{unique challenges} in data collection and design for the MIA that are different from other data domains such as computer vision. While an eavesdropper can observe a transmitted signal over the air, its received signal is different from (but potentially correlated with) the signal received by the target signal classifier due to different channel characteristics. Therefore, the data collected by the adversary is inherently different from the data input to the target signal classifier. For RF fingerprinting, the service provider runs its DL classifier to determine whether the received signal is from an authorized user or not. The input of the DL classifier is the I/Q data and the underlying decision process of user classification is based on the RF fingerprint of the user that depends on the radio device, waveform and channel characteristics of that particular user. In this paper, we consider a \emph{black-box MIA}, where the adversary does not know the target classifier. Even if the adversary knows this classifier (namely, the underlying DNN model), it may not use it to identify whether a signal is from an authorized user or not, since the signal received at the adversary is different from the signal received at the service provider. To overcome these challenges, the adversary builds a \emph{surrogate classifier} by using the overheard signals as the input. With this surrogate classifier, the adversary can launch the MIA to determine for its received signal, whether the corresponding signal received at the service provider has been used in the training data or not. 

We set up test scenarios of one service provider (such as a gNodeB in 5G or beyond applications) and some authorized users (such as user equipments (UEs) in terms of IoT devices). 
Signals of each user are transmitted over the air and thus are changed by both channel and device-specific phase shift and transmit power effects. The target DL classifier can reliably classify users (with close to $100$\% under various settings). On the other hand, an adversary observes spectrum data and classification results (by observing whether a user is accepted for communications) to build a surrogate classifier to classify its received signals. The adversary then launches the MIA to infer whether for a signal received at the adversary, its corresponding signal received at the service provider is a member of the training data or not. We consider two settings: (i) non-member signals (signals that are not in the training data set) can be generated by the same radio devices that generate member signals (signals that are in the training data set), or (ii) non-member signals are generated by other radio devices.
In the first setting, the accuracy of the MIA reaches $88.62$\% when the signal-to-noise-ratio (SNR) is high (at $10$ dB) while the accuracy of the MIA is $77.01$\% when the SNR is low (at $3$ dB). 
In the second setting, there are some radio devices that generate signals as training data.
These signals cover both strong and weak signals. There is also another device that generates signals as non-member data to test the MIA performance. The accuracy of the MIA reaches $97.88$\%.

Since wireless channels are random, they add uncertainties on received signals. Therefore, we study the impact of \emph{noisy variations} in received signals by changing the question on whether a particular received signal is in training data to the question on whether some its noisy variations are in training data. For that purpose, we generate multiple samples with different levels of noisy variations and use either the average or maximum score in the MIA when evaluating the attack success. If we use the average score, the accuracy of the MIA decreases with the level of noisy variations. If we use the maximum score, the accuracy of the MIA on member samples (authorized users) increases while the accuracy of the MIA on non-member samples (unauthorized users) decreases with the level of noisy variations.

We then develop a \emph{proactive defense} scheme for the MIA. The service provider first needs to build a \emph{shadow MIA model}. Then, it applies the defense using this shadow MIA model. For that purpose, perturbations (some controlled noise) are added in the classification process such that (i) there is no change made on classification results and (ii) the MIA in the presence of defense achieves low accuracy. We formulate this defense as an optimization problem and modify it to an unconstrained optimization by changing of variables and using loss function to remove constraints. We then apply gradient search to find the optimal perturbation. We show that this defense scheme effectively protects against the MIA launched by the adversary and reduces the accuracy from $97.88$\% to $50$\%.

The novel contributions of the paper are summarized as follows:
\begin{enumerate}
    \item We present the first MIA that is launched against a wireless classifier over the air to infer about training data and leak private information on waveform, device, and channel characteristics.
    \item  We consider two settings for the MIA: (i) the  MIA  should  be  able  to  identify  signals from the same radio device as member and non-member, and (ii)  non-member signals are generated by different radio devices.
    \item We extend the MIA such that it is launched by using not  only  received  signals  but  also  their  noisy  variations by accounting for channel variations.
    \item We show through detailed numerical results that the success of the MIA is high, i.e., the MIA can infer the training data membership of the wireless signal classifier with high accuracy.
    \item We present a defense scheme to protect wireless signal classifiers from the MIA and show that this defense can reduce the accuracy of the MIA significantly.
\end{enumerate}

The rest of the paper is organized as follows.
Section~\ref{sec:scenario} presents the system model.
Section~\ref{sec:classifier} describes the target classifier at the service provider.
Section~\ref{sec:mia} presents the MIA.
Section~\ref{sec:defense} presents the defense scheme for the MIA. Section~\ref{sec:result} presents the numerical results for the MIA and the defense.
Section~\ref{sec:conclusion} concludes the paper.

\section{System Model}
\label{sec:scenario}
 \begin{figure}
   \centering
   \includegraphics[width=0.8\columnwidth]{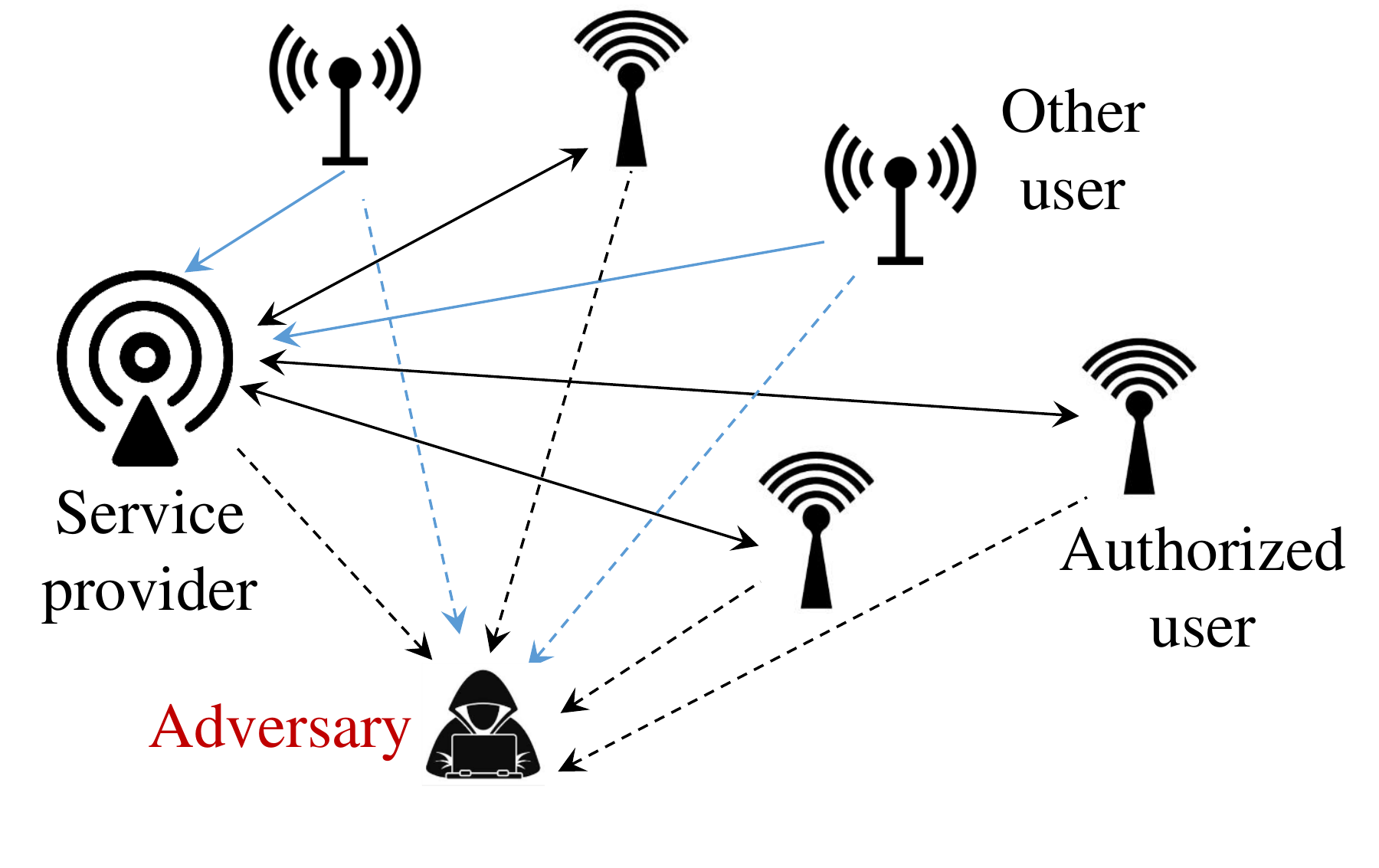}
   \caption{The MIA scenario with a service provider, an adversary, authorized users, and other users.}\label{fig:system}
 \end{figure}

We consider a wireless system that provides different services to users on the same physical network as shown in Fig.~\ref{fig:system}. Example use cases include, but are not limited to, network slices in 5G and beyond applications, and IoT networks. A wireless signal classifier is used for PHY-layer authentication to detect authorized users of these services based on their signal characteristics. An adversary aims to launch the MIA such that it can determine which users' signals are used in the training data. The adversary can then generate similar signals to gain access.

To authorize users for a particular service, the service provider uses a DL classifier, i.e., a DNN has been trained to classify users as authorized or not.
During the training process, each authorized user transmits some signals (that are labeled as `authorized' class) received subject to channel and noise effects, and other signals are labeled as `unauthorized' class.
The training data includes signals with different modulations, device-specific phase shifts as well as channel-specific gains and phase shift offsets in received signals. We assume that the environment does not change during training and test time beyond variations in channel realizations.
As we will later show in Section~\ref{sec:result}, such a classifier can achieve high accuracy and thus the service provider can reliably detect authorized users.

  \begin{figure}
   \centering
   \includegraphics[width=0.8\columnwidth]{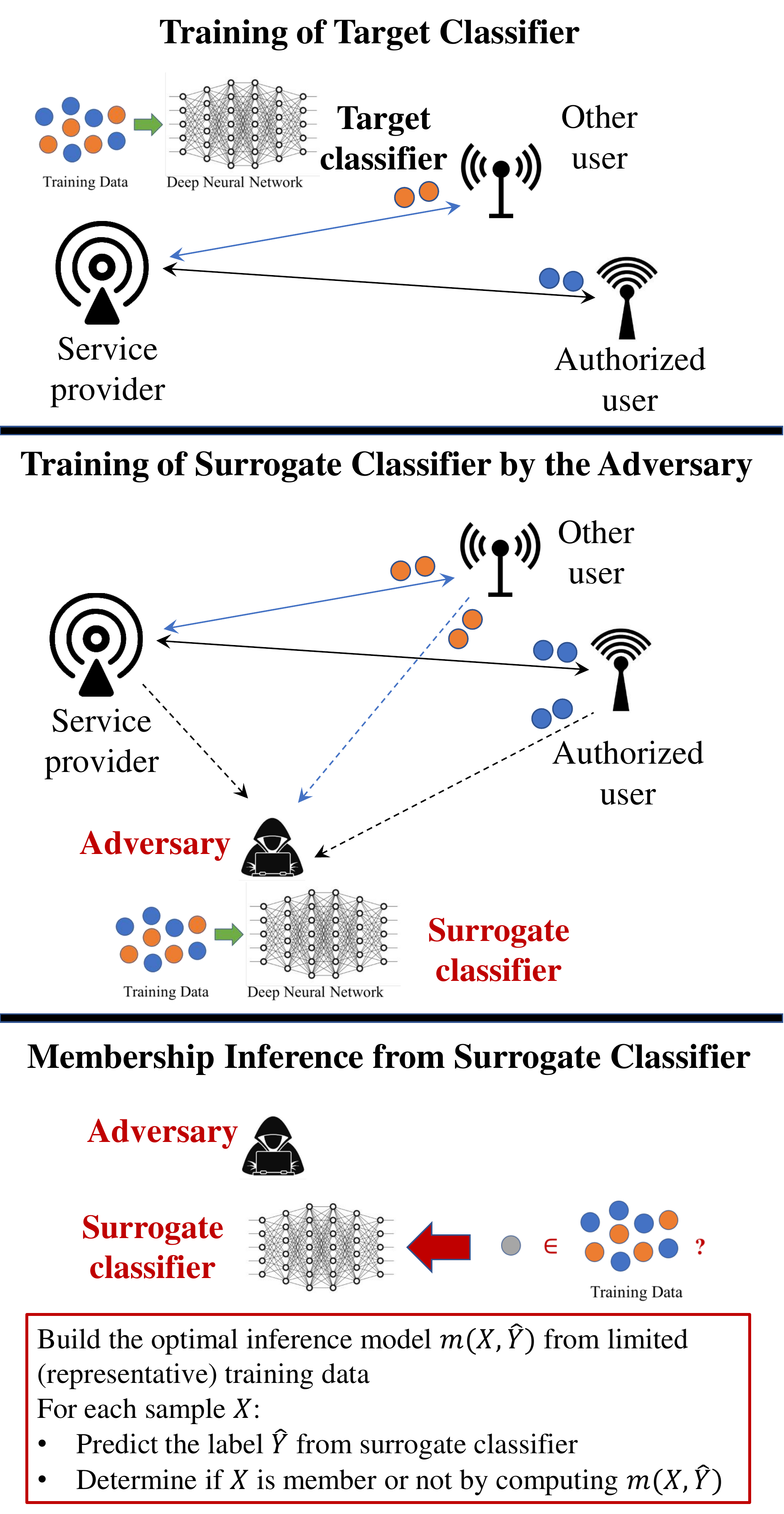}
   \caption{top: training of target model; middle: training of surrogate classifier by the adversary; bottom: membership inference attack from surrogate classifier.}\label{fig:procedure}
 \end{figure}
The process of launching the MIA is illustrated in Fig.~\ref{fig:procedure}. In this over-the-air MIA, an adversary collects signals transmitted from users and observes (overhears) classification results (their start of communications with the service provider as an indicator of service)
due to the shared and broadcast nature of the wireless medium.
Thus, the adversary can determine the class for each collected signal. The adversary can further build a surrogate classifier based on its over-the-air collected data to obtain classification labels on more signals. This corresponds to an exploratory (inference) attack \cite{Shi17HST}. The adversary further analyzes signals and the outputs of classification process
to launch the MIA that can determine whether a received signal is in the training data or not. Once the MIA is successful and signal characteristics of interest (e.g., device and channel) are identified, the adversary may perform other attacks, e.g., it may generate signals similar to those used in training data to gain access. The details of the MIA attack are provided in Sec.~\ref{sec:mia}.

  \begin{figure}
   \centering
   \includegraphics[width=0.8\columnwidth]{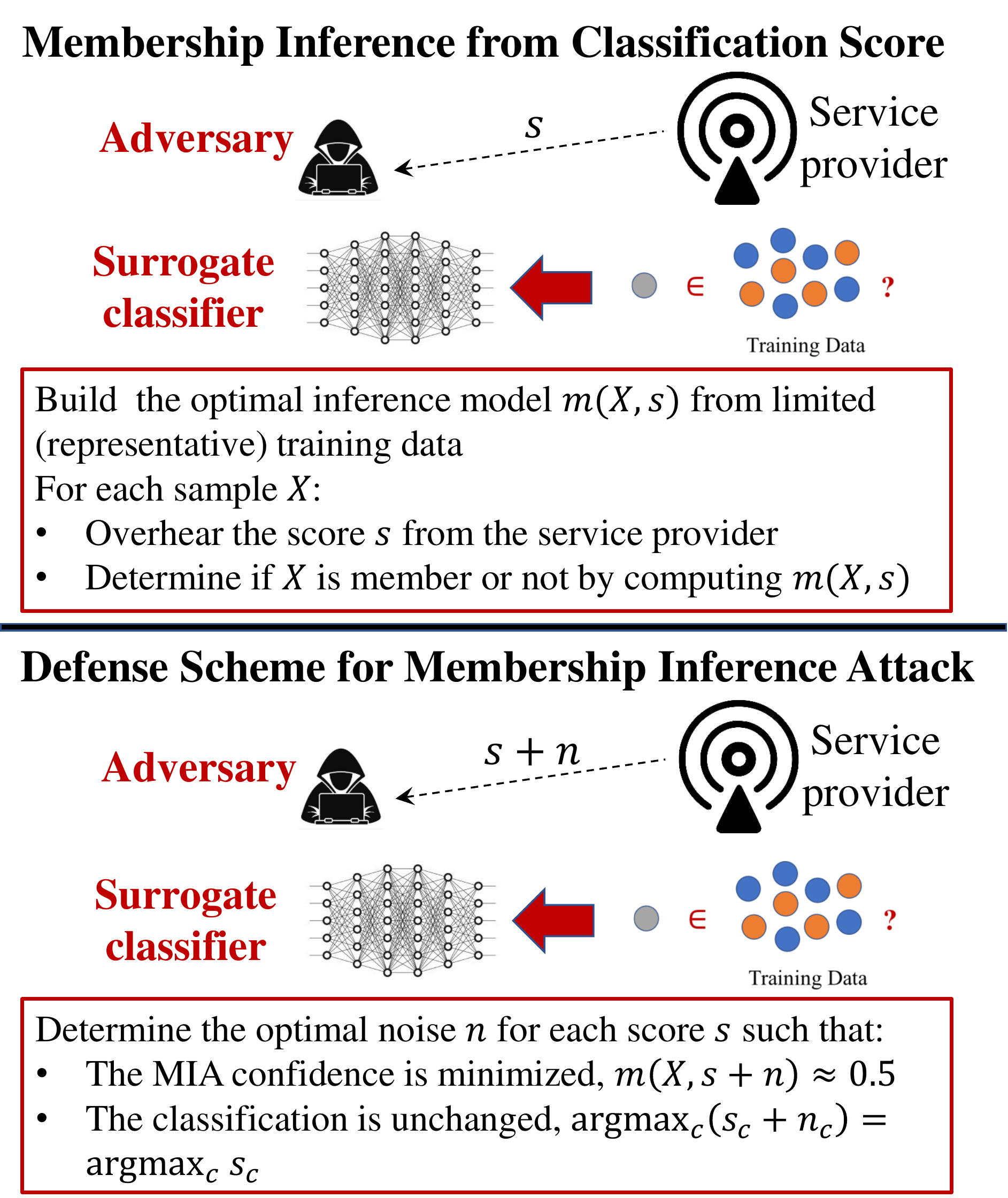}
   \caption{top: membership inference attack; bottom: defense scheme for membership inference.}\label{fig:mia-procedure}
 \end{figure}

The process of defending against the MIA is illustrated in Fig.~\ref{fig:mia-procedure}.
The service provider first builds a shadow MIA to model and analyze the potential MIA launched by the adversary. Then, the proactive defense analyzes and changes the classification outputs so that the shadow MIA, as well as MIA, cannot work well.
The details of the defense are provided in Sec.~\ref{sec:defense}.

\section{The Target Classifier at the Service Provider}
\label{sec:classifier}

We consider a service provider that aims to provide communications for some authorized users that use a particular modulation (in particular, QPSK) and has certain radio device and channel characteristics.
A simple classifier just for modulation recognition does not achieve this objective since it can only tell whether a signal is using QPSK or not, but cannot tell whether the source of a received signal is one of the authorized users as this decision also depends on the characteristics of radio devices and channels experienced by users.
To detect authorized users, we need to consider user-specific properties.
In particular, each user as a transmitter has its own phase shift due it unique radio hardware. Moreover, the channel from a user to the service provider has its own channel gain and phase shift. As a consequence, the phase shift and power of received signals are unique properties for users.
Denote $\phi_i$ and $\phi_{ib}$ as the phase shift of user $i$ and channel from $i$ to service provider $b$, respectively, and $g_{ib}$ as channel gain from $i$ to $b$. For example, if the raw data is two bits $00$ and transmit power is $p$, the received phase shift should be $\frac{\pi}{4} + \phi_i + \phi_{ib}$ and the received power should be $g_{ib} p$.
In reality, the collected data may have small random errors, i.e., $n_\phi$ for noise on phase shift and $n_p$ for noise on power.
Thus, a service provider collects phase shift and power of received signals as user-specific properties and uses them as features to build a classifier to detect authorized users. The labels of the classifier for users are `authorized' and `unauthorized'.

All authorized users use QPSK-modulated signals, whereas unauthorized users may either use QPSK-modulated or BPSK-modulated signals.
The service provider again collects the phase shift and the power as features. For example, if the raw data is a bit $0$, the received phase shift from transmitter $j$ is $\phi_j + \phi_{js} + n_\phi$ and the received power should be $g_{js} p + n_p$.
To have the same number of features, data is collected for each bit, i.e., data is collected twice for each coded QPSK symbol (corresponding two bits). Then, for $n$ bits, the number of collected features is $2n$, including $n$ phase shifts and $n$ powers.

During the training period, the service provider collects samples ($2n$ features and a label). 
Then, it trains a feedforward neural network as the DL classifier $C$. We consider three hidden layers for $C$, each with $100$ neurons. ReLU is used as the activation function at hidden layers, and softmax is used as the output layer. The deep neural network is trained with backpropagation algorithm and Adam optimizer using cross-entropy as the loss function. The process of building classifier $C$ is shown in the top portion of Fig.~\ref{fig:procedure}.

\section{Membership Inference Attack}
\label{sec:mia}

The goal of the MIA is to identify data samples that have been used to train a ML classifier (as studied in computer vision and other data domains \cite{Shokri17, Nasr18, Yeom18, Salem18, Leino19, Nasr19, Song19, Song19-2, Truex19, Jia20, Song20}).
One possible application of the MIA in the wireless domain is a privacy attack on PHY-layer signal authentication, where the adversary aims to identify the signal samples that have been used in the training of a wireless signal classifier. Then, the adversary can leverage these signal samples and leaked information on waveform, device and channel characteristics of authorized users to generate  signals in order to obtain service from the provider.
The training data and the general data usually have different distributions (e.g., due to differences of radios and channels in training and test times) and the classifier may overfit the training data. The MIA analyzes the overfitting to leak private information.

We first give an insight on how the MIA works. Suppose the general distribution (of signal samples) is $\theta^*$ and the training data distribution is $\hat \theta$. For a given data sample $x'$, we can determine the probability of $x'$ generated by $\theta^*$ (or by $\hat \theta$). If $P_{\hat \theta} (x')>P_{\theta^*} (x')$, then $x'$ is likely generated by $\hat \theta$ (where $P_{\theta}(x)$ denotes the probability of distribution $\theta$ evaluated at $x$. Moreover, we can calculate a confidence value for the MIA by $\frac{P_{\hat \theta} (x')}{P_{\hat \theta} (x')+P_{\theta^*} (x')}$. This analysis provides a prediction on whether any given data sample is in the training data or not, along with a confidence value for this prediction. Note that if $\theta^* = \hat \theta$, the confidence value is always $0.5$, i.e., we cannot make any better prediction than a blind guess. Thus, the basis of the MIA is the difference between $\theta^*$ and $\hat \theta$.
Next, we provide details of the MIA.

Suppose that each sample in the training data set is represented by a set of features $F$ and is labeled as one of two classes following a supervised learning approach. 
The MIA aims to identify whether a given sample is in the training data set to build the given classifier or not. This attack may be a white-box attack, i.e., the target classifier is available to the adversary, or a black-box attack, where the adversary does not know the classifier, but it can  collect data from the target classifier. In this paper, we consider a \emph{black-box attack}.
To launch an effective MIA, we consider a general approach as follows. Suppose that features include all (useful, but potentially biased and noisy) information, where useful information in $F_u$ can be used to identify the class, biased information $F_b$ is due to the different distributions of training data and general test data, and noisy information $F_n$ is other information with no statistical significance. Note that to simplify discussion, we assume each feature includes only one type of information. For the general case that one feature includes multiple types of information, we can divide it into multiple features to meet our assumption. DL is relied upon to extract useful and biased information while ignoring noisy information. Then, a classifier is optimized to fit on useful and biased information ($F_u$ and $F_b$). While fitting on $F_u$ can provide correct classification on general test data, fitting on $F_b$ corresponds to overfitting, which provides correct classification on the given training data but wrong classification on general test data.

For both white-box and black-box MIAs, overfitting is the key factor leading to privacy issues as the classifier (or the inferred classifier) memorizes some characteristics of the training data in $F_b$ and reflects it in the model's output behavior. Thus, we can infer the training data membership based on overfitting. In particular, if overfitting exists in training data, some features not related with a class may be used for classification. Therefore, a sample with such a feature is likely in the training data. By identifying such features in $F_b$, we can predict the membership and also provide a confidence score on such predictions. However, the distribution of training data may not be available and the distribution of general test data is unknown. Thus, features related with overfitting cannot be obtained directly for membership inference. Since DL models are sensitive to training data, the adversary can investigate parameters in the target (or inferred) classifier based on local linear approximation for each layer and the combination of all layers, as studied in \cite{Nasr18}. This approach builds a classifier for membership inference. Unlike the naive attack, where only an inference result is obtained, the developed classifier can also provide a confidence score on the inferred results.

MIA can be launched using the sensed signals for samples with either class or score information \cite{Nasr18,Jia20}.
We first present how to launch MIA based on data sample $X$ and class $\hat Y$.
The adversary can determine class $\hat Y$ for each $X$ from its score $s$.
Alternatively, the adversary can first build a surrogate classifier, and then use this surrogate classifier $\hat C$ to determine a class $\hat Y$ for any given data sample $X$.
Note that classifiers $\hat C$ and $C$ are not equal since for the same transmitted signal, the received signals at the service provider and at the adversary are different, i.e., the inputs to $\hat C$ and $C$ are different.
Instead, these two classifiers should provide the same label for their inputs on the same transmitted signal.
The process of building classifier $\hat C$ is shown in the middle portion of Fig.~\ref{fig:procedure}.

The MIA requires the adversary to further build an inference model $m(X, \hat Y)$ to provide a probability of training data for any sample and its label.
Let $P_{\mathcal D}(X, \hat Y)$ and $P_{\bar {\mathcal D}}(X, \hat Y)$ denote the conditional probabilities of $(X, \hat Y)$ for samples in training data ${\mathcal D}$ or not, respectively.
Then, the gain function the for MIA developed in \cite{Nasr18} is given by
\begin{eqnarray} \label{eq:gain}
G(m) &=& \frac{1}{2} E_{(X, \hat Y) \sim P_{\mathcal D}(X, \hat Y)} [\log m(X, \hat Y)] \\
&& + \frac{1}{2} E _{(X, \hat Y) \sim P_{\bar {\mathcal D}}(X, \hat Y)} [\log (1- m(X, \hat Y))] \; \nonumber,
\end{eqnarray}
where $E[\cdot]$ is the expectation function. We use weight $\frac{1}{2}$ because we want to maximize the gain on both samples that are in training data and not in training data.
In reality, we do not have $P_{\mathcal D}(X, \hat Y)$ and $P_{\bar {\mathcal D}}(X, \hat Y)$, and thus cannot calculate the gain defined in (\ref{eq:gain}).
Thus, we consider an empirical gain on a data set ${\mathcal D}^A$, which is a representative subset of ${\mathcal D}$, and a data set ${\bar {\mathcal D}}^A$, which is a representative subset of $\bar {\mathcal D}$.
The empirical gain is defined in \cite{Nasr18} as
\begin{eqnarray}
G_{{\mathcal D}^A, {\bar {\mathcal D}}^A} (m) &=& \frac{1}{2 |{\mathcal D}^A|} E_{(X, \hat Y) \in {\mathcal D}^A} [\log m(X, \hat Y)] \\
&& + \frac{1}{2 |{\bar {\mathcal D}}^A|} E_{(X, \hat Y) \in {\bar {\mathcal D}}^A} [\log (1- m(X, \hat Y))]. \nonumber
\label{eq:gain2}
\end{eqnarray}
To find the optimal inference model $m(X, \hat Y)$, we need to solve
\begin{eqnarray}
\max_m G_{{\mathcal D}^A, {\bar {\mathcal D}}^A} (m). \label{eq:opt}
\end{eqnarray}
We can again see the need of different distributions of training data and general data from (\ref{eq:opt}).
If there is no difference, two representative subsets can be the same, i.e., ${\mathcal D}^A = {\bar {\mathcal D}}^A$.
For such sets, the optimal solution to the above problem is $m(X, \hat Y) = 0.5$ for all samples, i.e., the MIA is not successful if there is no difference on distributions.
The process of launching the MIA based on the surrogate classifier is shown in the bottom portion of Fig.~\ref{fig:procedure}.

For the MIA, we consider two settings. In the first setting, non-member signals can be generated by the same radio devices that generate member signal. In the second setting,  non-member signals are generated by other radio devices. For the second setting,
the service provider sends the classification score to each user, which shows the confidence of classification. Note that only PHY-layer authentication is considered for the UEs such as IoT devices without relying on encryption that may be computationally difficult for simple UE devices. This score serves as CTS for authorized users.
If a user is not classified as an authorized user, the service provider does not grant it any access.
If an authorized user is classified as non-authorized (which occurs with low probability), it can report this error along with the score and can reconfigure its signal characteristics (e.g., by changing power or location) to improve its score and gain access later.
We then present how to launch the MIA based on data sample $X$ and score $s$.
The MIA requires the adversary to further build an inference model $m(X, s)$ to provide a probability of training data for any sample and its score.
We again consider a data set ${\mathcal D}^A$ as a representative subset of ${\mathcal D}$ and a data set ${\bar {\mathcal D}}^A$ as a representative subset of $\bar {\mathcal D}$.
We build the inference model on ${\mathcal D}^A$ and ${\bar {\mathcal D}}^A$ to maximize the inference accuracy, i.e., the ratio of $m(X, s) = 1$ for $X \in {\mathcal D}^A$ and $m(X, s) = 0$ for $X \in {\bar {\mathcal D}}^A$ over all samples in ${\mathcal D}^A$ and ${\bar {\mathcal D}}^A$.
The process of launching the MIA using classification score is shown in the top portion of Fig.~\ref{fig:mia-procedure}.

\section{Defense Scheme}
\label{sec:defense}

To develop the defense, the defender (e.g., the gNodeB) first builds a shadow MIA and then develops its proactive defense against this shadow MIA. For that purpose, a shadow MIA is trained on a set of member samples and a set of non-member samples as we discussed in the last section. We then test the performance of this shadow MIA on some member and non-member samples.
If $\hat m(X,s)=1$ for a sample $X$ and its score $s$, we claim that $X$ is a member, otherwise we claim $X$ as a non-member.
One observation is that for a set of scores, there are many possible orders of scores to build a vector, which corresponds to many inputs. The MIA problem on such a large space is not easy to train. Instead, we sort these scores to reduce the size of the input space and find that the performance of MIA can be improved by sorting.

Then, the defense scheme against this shadow MIA is developed as follows.
The main idea of the defense is adding some perturbation (noise) vector $n$ to $s$ such that (i) the shadow MIA accuracy is low and (ii) the classification results are still good.
\begin{itemize}
\item For (i), we aim to make $\hat m(X,s+n) \approx 0.5$, which means that the shadow MIA cannot identify a member/non-member with high confidence. We can formulate the objective as
    $$\min_n |\hat m(X,s+n) -0.5| \; .$$
    Note that it is not a good strategy to push $\hat m(X,s+n)$ close to $0$ when $\hat m(X,s) > 0.5$ and to push $\hat m(X,s+n)$ close to $1$ when $\hat m(X,s) < 0.5$. This is because that under such a defense scheme, one can switch the MIA decisions to obtain good MIA results.

\item Since the classification result is the class with the largest score, for (ii) we aim to achieve \begin{eqnarray}
    \argmax_c \{ s_c+n_c \} = c^* \; , \label{eq:class}
    \end{eqnarray}
    where $s_c$ and $n_c$ are the $c$-th element in $s$ and $n$, respectively, $c^* = \argmax_c s_c$ is the predicted class by the target classifier $C$.
    Note that we aim to keep the same results as the classification results by $C$, not achieving the true class, since the true class information is not available.
\end{itemize}
The classification scores should be non-negative and sum up to $1$. Then, we have the following constraints.
\begin{eqnarray}
s_c+n_c \ge 0 \; , \label{eq:score1} \\
\sum_c (s_c+n_c) = 1 \; . \label{eq:score2}
\end{eqnarray}
Therefore, for any score $s$ (with $s_c \ge 0$ and $\sum_c s_c =1$), we solve the following problem to determine the optimal perturbation $n$.
\begin{equation} \label{eq:opt1}
\begin{split}
\min & \hspace{0.5cm} |\hat m(X,s+n) -0.5| \\
\mbox{subject to} & \hspace{1cm}  (\ref{eq:class}), (\ref{eq:score1}), (\ref{eq:score2}) \; .
\end{split}
\end{equation}

We can simplify this optimization problem (\ref{eq:opt1}) by changing variables.
Define $\tilde z_c = \log(s_c + n_c)$. 
Then, we can use the softmax function $\sigma(\tilde z)$ to calculate $s + n$ from $\tilde z$. Moreover, constraints (\ref{eq:score1}) and (\ref{eq:score2}) will always be met and thus can be omitted. Since $\log(\cdot)$ is a monotone increasing function, constraint (\ref{eq:class}) can be replaced by
\begin{eqnarray}
\argmax_c \tilde z_c = c^* \; . \label{eq:class2}
\end{eqnarray}
Then, we end up with a simpler problem as follows.
\begin{equation} \label{eq:opt2}
\begin{split}
\min & \hspace{0.5cm} |\hat m(X,\sigma(\tilde z)) -0.5| \\
\mbox{subject to} & \hspace{1.5cm} (\ref{eq:class2}) \; ,
\end{split}
\end{equation}
where $\sigma(\tilde z)_j = \frac{e^{\tilde z_j}}{\sum_j e^{\tilde z_j}}$.
We can reformulate the problem (\ref{eq:opt2}) as an unconstrained optimization by defining the following loss term for (\ref{eq:class2}).
\begin{eqnarray}
L = \max\{ \max_{c \ne c^*} \tilde z_c - \tilde z_{c^*}, 0\} \; . \label{eq:loss}
\end{eqnarray}
Then, the optimization problem is changed to
\begin{eqnarray}
\min & |\hat m(X,\sigma(\tilde z)) -0.5| + \lambda L \; , \label{eq:opt}
\end{eqnarray}
where $\lambda > 0$ is a constant factor for the loss term.
We can start with $\tilde z_c = \log s_c$ and apply the gradient search algorithm to find the optimal solution for the above unconstrained problem with a given $\lambda$.
We can try different $\lambda$ values to find the best solution.
But there is no need to find an optimal solution for defense purpose, i.e., we can stop as long as we find a solution with $\hat m(X,\sigma(\tilde z)) \approx 0.5$ and $L =0$.

Finally, recall that the shadow MIA is working on a sorted score vector. An additional defense step is keeping the position $c^*$ of the largest score unchanged and shuffling other scores. This does not change the performance of the shadow MIA but will make further protection for the MIA.
The defense process for the MIA is shown in the bottom portion of Fig.~\ref{fig:mia-procedure}.

\section{Performance Evaluation of the MIA and the Defense}
\label{sec:result}

To evaluate the performance of the MIA and the defense, we consider one service provider, some authorized users, and other unauthorized users. Authorized users transmit QPSK-modulated signals while other users can transmit with either BPSK-modulated or QPSK-modulated signals.
The service provider aims to distinguish signals from authorized and other users by using its classifier $C$. This is not a simple modulation classifier since both authorized and unauthorized users may transmit signals with the same modulation.
Each sample has phase shift and power values for $16$ bits, i.e., there are $32$ features.
These values are collected with noise within small bounds $[-e_u, e_u]$, where $e_u$ for phase values is $0.1$ and for power values is the same as noise.

The adversary can overhear signals and outputs of classification process, and identify who gains access. 
Then, it can launch the MIA. The service provider can make a noisy decision such that the MIA by the adversary is not successful.
For this purpose, the service provider builds a shadow MIA and determine the optimal perturbation on this shadow MIA. We will show that the determined perturbation also works for the MIA by the adversary.

\subsection{The MIA Performance}
\label{subsec:mia}

Given that a user device generates some signals as training data, one important question on the MIA is whether to consider other signals generated by this device as non-member or member in the MIA. By definition, other signals should be considered as non-member since they are not used as training data. However, signals generated by the same device share many common properties (i.e., fingerprint). Therefore, identifying them as non-member is challenging. In this section, we will study the MIA in two different settings. In the first setting, we require that the MIA should be able to identify signals from the same device as member and non-member while in the second setting, non-member signals are generated by different devices.

\subsubsection{Setting 1}
We consider three authorized users.
The training data of $C$ has $8000$ signal samples.
Half of them are class $1$ samples, which correspond to QPSK signals from authorized users.
The remaining samples are BPSK signals from other users.
The training data set of $\hat C$ has $1000$ samples, one half is class $1$ data and the other half is class $0$ data. Note that the training data set of $\hat C$ is smaller than that of $C$, as the adversary may not have access or time to collect as many training data samples as in the target classifier $C$ that was trained before the attack.
In test time, we use $10000$ samples for $C$ and $1000$ samples for $\hat C$.
The surrogate classifier $\hat C$ achieves almost $100$\% accuracy. 

To evaluate the MIA performance, $1000$ samples from $8000$ training samples are used as member samples. Note that we do not use the received signals at the service provider since the adversary does not have access to the received signals at the adversary. Instead, we use corresponding signals received at the adversary.
In addition, $1000$ samples are used as non-member samples. Among these samples, half are QPSK signals from authorized users (class $1$ data) and half are QPSK from other users (class $0$) data.
We first consider the case that the signals of authorized users are stronger than the signals of other signals.
The SNR values are about $10$~dB and $3$~dB, respectively.
This MIA attack uses $32$ features on both phase shift and power, and the class of a sample as its input.
The accuracy of the MIA (i.e., the average accuracy of predicting member and non-member samples) is $88.62$\% and the confusion matrix is given in Table~\ref{table:test1}.

\begin{table}
\caption{Confusion matrix when authorized users have stronger signals.}
\centering
{\small
\begin{tabular}{c|c|c}
Real $\backslash$ Predicted & non-member & member \\ \hline \hline
non-member & 0.9152 & 0.0848 \\ \hline
member & 0.1429 & 0.8571 \\ \hline
\end{tabular}
}
\label{table:test1}
\end{table}

We also consider the scenario where the authorized users have weaker signals than other users, i.e., $3$~dB for authorized users and $10$~dB for other users.
The accuracy of the MIA is measured as $77.01$\% and the confusion matrix is given in Table~\ref{table:test4}.
These results show that as the authorized user signals become weaker, the success of the MIA drops.

\begin{table}
\caption{Confusion matrix when authorized users have weaker signals.}
\centering
{\small
\begin{tabular}{c|c|c}
Real $\backslash$ Predicted & non-member & member \\ \hline \hline
non-member & 0.9129 & 0.0871 \\ \hline
member & 0.3728 & 0.6272 \\ \hline
\end{tabular}
}
\label{table:test4}
\end{table}

Since there are random channel effects on received signals, we also consider the case when the MIA is launched by using not only received signals but also their noisy variations. The goal of the MIA is changed to determining whether both a given received signal and its noisy variations are in the training data or not.
We determine the range of phase and power feature values.
The level of noisy variations on a particular feature is measured by the ratio between changes and the maximum possible value for this feature.
For the obtained vector of the level of noisy variations on each feature, we define the total level of variation as the norm of this vector.
To study how considering noisy variations of received signals impacts the MIA, we generate different samples based on an original sample with $10$ different levels of noisy variations, changing from $0.1$ to $0.9$.
We generate $10$ samples at each level.
Since each sample has different classification scores, we use the average score or the maximum score to predict the original sample, and show results in Tables~\ref{table:avg} and \ref{table:max}, respectively. If the average score is used, the accuracy to predict member signals decreases with the level of noisy variations while the accuracy on non-member signals does not change much.
If the maximum score is used, the accuracy to predict member signals increases to $1$ even when small noisy variations are added while the accuracy on non-member signals decreases sharply with the level of noisy variations.

\begin{table}
\caption{Accuracy results when the average score is used for the level of noisy variations added to received signals.}
\centering
{\small
\begin{tabular}{c|c|c}
Noisy variation level & non-member & member \\ \hline \hline
0 & 0.9152 & 0.8571 \\ \hline
0.1 & 0.8594 & 0.5413 \\ \hline
0.2 & 0.9029 & 0.4688 \\ \hline
0.3 & 0.9241 & 0.4085 \\ \hline
0.4 & 0.9362 & 0.4007 \\ \hline
0.5 & 0.9118 & 0.3806 \\ \hline
0.6 & 0.9185 & 0.4163 \\ \hline
0.7 & 0.8940 & 0.3717 \\ \hline
0.8 & 0.8973 & 0.3884 \\ \hline
0.9 & 0.8783 & 0.4163 \\ \hline
\end{tabular}
}
\label{table:avg}
\end{table}

\begin{table}
\caption{Accuracy results when the maximum score is used for the level of noisy variations added to received signals.}
\centering
{\small
\begin{tabular}{c|c|c}
Noisy variation level & non-member & member \\ \hline \hline
0 & 0.9152 & 0.8571 \\ \hline
0.1 & 0.4252 & 1.0 \\ \hline
0.2 & 0.4040 & 1.0 \\ \hline
0.3 & 0.3761 & 1.0 \\ \hline
0.4 & 0.3270 & 1.0 \\ \hline
0.5 & 0.2478 & 1.0 \\ \hline
0.6 & 0.1931 & 1.0 \\ \hline
0.7 & 0.1741 & 1.0 \\ \hline
0.8 & 0.1563 & 1.0 \\ \hline
0.9 & 0.1350 & 1.0 \\ \hline
\end{tabular}
}
\label{table:max}
\end{table}

\subsubsection{Setting 2}

In the second setting, to build the target classifier, we use $10$ authorized devices to generate QPSK signals and 10 other devices to generate BPSK signals (i.e., $20$ classes). These classes cover both strong and weak signals. To have non-member signals, we put another device at a different location to generate QPSK and BPSK signals. We train the target classifier $C$ by a set $\mathcal A$ of $2000$ samples and test it on another set $\mathcal B$ of $2000$ samples.
This target classifier achieves high accuracy, namely $92.85$\%, for the $20$-class problem.

We now consider an adversary, which can collect scores by overhearing the output of the target classifier and wants to launch the MIA. Suppose that it overhears the scores of some samples with both members and non-members, i.e., scores of $2000$ samples in $\mathcal A_1+ \mathcal D_1$ for member set $\mathcal A_1 \in \mathcal A, |\mathcal A_1|=1000$ and non-member set $\mathcal D_1 \in \mathcal D, |\mathcal D_1|=1000$, to build its MIA classifier $m(\cdot)$, which maps a sample $X$ with its score $s$ to either $m(s)=1$ (member) or $m(s)=0$ (non-member). Then, this MIA classifier is used for the score of $2000$ samples in $\mathcal A- \mathcal A_1 + \mathcal D- \mathcal D_1$ to test its performance. The MIA achieves high accuracy $97.88$\% and its confusion matrix is shown in Table~\ref{table:mia-confusion}.

\begin{table}
\caption{Confusion matrix for the MIA.}
\centering
{\small
\begin{tabular}{c|c|c}
Real $\backslash$ Predicted & non-member & member \\ \hline \hline
non-member & 0.9905 & 0.0095 \\ \hline
member & 0.0330 & 0.9670 \\ \hline
\end{tabular}
}
\label{table:mia-confusion}
\end{table}

\subsection{The Defense Performance}
In this section, we mainly focus on the defense results for the second setting, since the first setting is very challenging from the MIA point of view and therefore the MIA is not necessarily always very effective even no defense is applied, e.g., the MIA accuracy is not high if signal strength is not strong (e.g., the accuracy to identify member is $62.72$\% in Table~\ref{table:test4}). The defense scheme cannot achieve much effect on this MIA, e.g., the reduction on accuracy is only about $5$\%.

To develop a defense scheme for the MIA, the service provider first builds a shadow MIA. For the second setting, we train a shadow MIA on set $\mathcal A$ for member samples and set $\mathcal C$ of $2000$ non-member samples. We then test the performance of this shadow MIA on set $\mathcal A$ and another set $\mathcal D$ of $2000$ non-member samples.
This shadow MIA is very successful with accuracy $99.00$\% and its confusion matrix is shown in Table~\ref{table:shadow-mia-confusion}.

\begin{table}
\caption{Confusion matrix for the shadow MIA.}
\centering
{\small
\begin{tabular}{c|c|c}
Real $\backslash$ Predicted & non-member & member \\ \hline \hline
non-member & 0.9895 & 0.0105 \\ \hline
member & 0.0095 & 0.9905 \\ \hline
\end{tabular}
}
\label{table:shadow-mia-confusion}
\end{table}

We start with $\tilde z_c = \log s_c$, $\lambda =10$, and apply the gradient search algorithm to solve (\ref{eq:opt}).
We try different $\lambda$ values to find optimal solution and stop once we find a solution with $\hat m(\sigma(\tilde z)) \approx 0.5$ and $L =0$.
With noise $n$, the accuracy of the shadow MIA reduces to $62.00$\% and the confusion matrix is shown in Table~\ref{table:shadow-mia-confusion-defense}. Hence, the defense is effective for the shadow MIA.

\begin{table}
\caption{Confusion matrix for the shadow MIA under defense.}
\centering
{\small
\begin{tabular}{c|c|c}
Real $\backslash$ Predicted & non-member & member \\ \hline \hline
non-member & 0.4125 & 0.5875 \\ \hline
member & 0.1725 & 0.8275 \\ \hline
\end{tabular}
}
\label{table:shadow-mia-confusion-defense}
\end{table}

We further check the performance of the MIA (developed in Section~\ref{subsec:mia}) under defense. The accuracy of the MIA is $50.00$\% and the confusion matrix is shown in Table~\ref{table:mia-confusion-defense}. This MIA claims all samples as members in training data under defense, which does not work at all.
A summary of the MIA performance with and without defense is given in Table~\ref{table:mia-defense}.

\begin{table}
\caption{Confusion matrix for the MIA under defense.}
\centering
{\small
\begin{tabular}{c|c|c}
Real $\backslash$ Predicted & non-member & member \\ \hline \hline
non-member & 0.0000 & 1.0000 \\ \hline
member & 0.0000 & 1.0000 \\ \hline
\end{tabular}
}
\label{table:mia-confusion-defense}
\end{table}

\begin{table}
\caption{The MIA accuracy under defense.}
\centering
{\small
\begin{tabular}{c|c|c}
 & Shadow MIA & MIA \\ \hline \hline
no defense & $99.00$\% & $97.88$\% \\ \hline
with defense & $62.00$\% & $50.00$\% \\ \hline
\end{tabular}
}
\label{table:mia-defense}
\end{table}

\section{Conclusion}
\label{sec:conclusion}

In this paper, we studied the MIA as a novel privacy threat against ML-based wireless applications. The target application is a DL-based classifier to identify authorized users by their RF fingerprint. An example use case for this attack is PHY-layer user authentication in 5G or IoT systems. The input of this model consists of the received power and the phase shift. An adversary launches the MIA to infer whether signals of interest have been used to train this wireless signal classifier or not. In this attack, the adversary needs to collect signals and their classification results by observing the spectrum. Then, it can build a surrogate classifier namely a functionally equivalent classifier as the target classifier at the intended receiver, e.g., a service provider. We showed that the surrogate classifier can be reliably built by the adversary under various settings. Then, the adversary launches the MIA to identify whether for a received signal, its corresponding signal received at the service provider is in the training data or not.

In the first setting where non-member signals can be generated by the same devices, the MIA accuracy is $88.62$\% for strong signals and $77.01$\% for weak signals.
We studied the case that the member inference is investigated not only for received signals but also their noisy variations due to random channel effects. If the average score is used to predict the membership inference for original signals and their noisy variations, the accuracy of the MIA decreases with the level of noisy variations. On the other hand, if the maximum score is used, the accuracy on member samples increases while the accuracy on non-member samples decreases. In the second setting where non-member signals are generated by different devices, the MIA achieves better performance ($97.88$\% accuracy). All these results indicate the MIA as a genuine threat for wireless privacy and show how the MIA can be effectively launched to infer private information from ML-based wireless systems over the air.

We further developed a defense scheme at the service provider that adds carefully crafted perturbations in the classification process such that there is no change on classification result but the MIA cannot work well. For the first setting, the MIA accuracy is originally not high and it is reduced by the defense only to a small extent (about $5$\%) while the defense is highly effective for the second setting and reduces the MIA accuracy to $50$\%.


\begin{thebibliography}{99}
	
\bibitem{WiseML2020}
Y. Shi, K. Davaslioglu, and Y. E. Sagduyu, ``Over-the-Air Membership Inference Attacks as Privacy Threats for Deep Learning-based Wireless Signal Classifiers," \emph{ACM Conference on Security and Privacy in Wireless and Mobile Networks (WiSec) Workshop on Wireless Security and Machine Learning (WiseML)}, 2020.

\bibitem{bc}
T. Erpek, T. O'Shea, Y. E. Sagduyu, Y. Shi, and T. C. Clancy, ``Deep Learning for Wireless Communications" in \emph{Development and Analysis of Deep Learning Architectures}, Springer, 2020

\bibitem{aml}
Y. E. Sagduyu, Y. Shi, T. Erpek, W. Headley, B.Flowers, G. Stantchev, and Z. Lu, ``When Wireless Security Meets Machine Learning: Motivation, Challenges, and Research Directions," \emph{arXiv preprint} arXiv:2001.08883, 2020/

\bibitem{aml2}
D. Adesina D, C. C. Hsieh, Y. E. Sagduyu, and L. Qian, ``Adversarial Machine Learning in Wireless Communications using RF Data: A Review," \emph{arXiv preprint} arXiv:2012.14392, 2020.

\bibitem{Shi2018}
Y. Shi, Y. E Sagduyu, T. Erpek, K. Davaslioglu, Z. Lu, and J. Li,
``Adversarial Deep Learning for Cognitive Radio Security: Jamming Attack and Defense Strategies," \emph{IEEE International Conference on Communications (ICC) Workshop on Promises and Challenges of Machine Learning in Communication Networks}, 2018.

\bibitem{terpek}
T. Erpek, Y. E. Sagduyu, and Y. Shi, ``Deep Learning for Launching and Mitigating Wireless Jamming Attacks," \emph{IEEE Transactions on Cognitive Communications and Networking}, Mar. 2019.

\bibitem{NetSliceAttack}
Y. Shi, Y. E. Sagduyu, T. Erpek, and M. C. Gursoy, ``How to Attack and Defend 5G Radio Access Network Slicing with Reinforcement Learning," \emph{arXiv preprint} arXiv:2101.05768, 2021.

\bibitem{Larsson}
M.  Sadeghi  and  E.  G.  Larsson, ``Adversarial  Attacks  on  Deep-learning based Radio Signal Classification," \emph{IEEE Communications Letters}, Feb. 2019.

\bibitem{Larsson2}
M. Sadeghi and E. G. Larsson, ``Physical Adversarial Attacks Against End-to-end Autoencoder Communication Systems," \emph{IEEE Communications Letters}, May 2019.

\bibitem{Kim}
B. Kim, Y. E. Sagduyu, K. Davaslioglu, T. Erpek, and S. Ulukus, ``Over-the-Air Adversarial Attacks on Deep Learning Based Modulation Classifier over Wireless Channels," \emph{Conference on Information Sciences and Systems (CISS)}, 2020.

\bibitem{Kim2}
B. Kim, Y. E. Sagduyu, K. Davaslioglu, T. Erpek, and S. Ulukus, ``Channel-Aware Adversarial Attacks Against Deep Learning-Based Wireless Signal Classifiers," \emph{arXiv preprint} arXiv:2005.05321.

\bibitem{Mao2021}
Y. Lin, H. Zhao, Y. Tu, S. Mao, and Z. Dou, ``Threats of Adversarial Attacks in DNN based Modulation Recognition," \emph{IEEE INFOCOM}, 2020.

\bibitem{Kim4}
B. Kim, Y. E. Sagduyu, K. Davaslioglu, T. Erpek, and S. Ulukus, ``Adversarial Attacks with Multiple Antennas against Deep Learning-based Modulation Classifiers," \emph{IEEE Global Communications Conference (GLOBECOM)}, 2020.

\bibitem{Kim5}
B. Kim, Y. E. Sagduyu, T. Erpek, K. Davaslioglu, and S. Ulukus, ``Channel Effects on Surrogate Models of Adversarial Attacks against Wireless Signal Classifiers," IEEE International Conference on Communications (ICC), 2021.

\bibitem{Larsson3}
B. Manoj, M. Sadeghi, and E. G. Larsson, ``Adversarial Attacks on Deep Learning based Power Allocation in a Massive MIMO Network," \emph{arXiv preprint} arXiv:2101.12090, 2021.

\bibitem{Kim6}
B. Kim, Y. E. Sagduyu, T. Erpek, and S. Ulukus, ``Adversarial Attacks on Deep Learning Based mmWave Beam Prediction in 5G and Beyond," \emph{IEEE Statistical Signal Processing Workshop}, 2021.

\bibitem{sahay2021deep}
R. Sahay, C. G. Brinton, and D. J. Love, ``Ensemble-based Wireless Receiver Architecture for Mitigating Adversarial Interference in Automatic Modulation Classification," \emph{arXiv preprint arXiv:2104.03494}, 2021.

\bibitem{Bahramali}
A. Bahramali, M. Nasr, A. Houmansadr, D. Goeckel, and D. Towsley, ``Robust Adversarial Attacks Against DNN-Based Wireless Communication Systems," \emph{arXiv preprint arXiv:2102.00918}, 2021.

\bibitem{Jinho}
J. Yi, Jinho and A. El Gamal, ``Gradient-based Adversarial Deep Modulation Classification with Data-driven Subsampling, \emph{arXiv preprint arXiv:2104.06375}, 2021.

\bibitem{Yalin2019}
Y. E. Sagduyu, Y. Shi, and T. Erpek, "IoT Network Security from the Perspective of Adversarial Deep Learning," \emph{IEEE International Conference on Sensing, Communication and Networking (SECON) Workshop on Machine Learning for Communication and Networking in IoT}, 2019.

\bibitem{Headley2019}
S. Bair, M. DelVecchio, B. Flowers, A. J. Michaels, and W. C. Headley, ``On the Limitations of Targeted Adversarial Evasion Attacks against Deep Learning Enabled Modulation Recognition," \emph{ACM Workshop on Wireless Security and Machine Learning (WiseML)}, 2019.

\bibitem{Headley2019-1}
B. Flowers, R. M. Buehrer, and W. C. Headley, ``Evaluating Adversarial Evasion Attacks in the Context of Wireless Communications," \emph{IEEE
Transactions on Information Forensics and Security}, 2020.

\bibitem{Headley2020}
M. DelVecchio, V. Arndorfer, and W. C. Headley, ``Investigating a Spectral Deception Loss Metric for Training Machine Learning-based Evasion Attacks," \emph{ACM Workshop on Wireless Security and Machine Learning}, 2020.

\bibitem{Silvija2019}
S. Kokalj-Filipovic and R. Miller, ``Adversarial Examples in RF Deep Learning: Detection of the Attack and its Physical Robustness," \emph{IEEE Global Conference on Signal and Information Processing (GlobalSIP)}, 2019.

\bibitem{Silvija2019-2}
S. Kokalj-Filipovic, R. Miller, and J. Morman, ``Targeted Adversarial Examples against RF Deep Classifiers," \emph{ACM Workshop on Wireless Security and Machine Learning (WiseML)}, 2019.

\bibitem{Silvija2019-3}
S. Kokalj-Filipovic, R. Miller, N. Chang, and C. L. Lau, ``Mitigation of Adversarial Examples in RF Deep Classifiers Utilizing Autoencoder Pre-training," \emph{International Conference on Military Communications and Information Systems (ICMCIS)}, 2019.

\bibitem{Sagduyu1}
Y. E. Sagduyu, Y. Shi, and T. Erpek, ``Adversarial Deep Learning for Over-the-Air Spectrum Poisoning Attacks," \emph{IEEE Transactions on Mobile Computing}, Feb. 2021.

\bibitem{YiMilcom2018}
Y. Shi, T. Erpek, Y. E Sagduyu, and J. Li, ``Spectrum Data Poisoning with Adversarial Deep Learning," \emph{IEEE Military Communications Conference (MILCOM)}, 2018.

\bibitem{Luo2019}
Z. Luo, S. Zhao, Z. Lu, J. Xu, and Y. E. Sagduyu, ``When Attackers Meet AI: Learning-empowered Attacks in Cooperative Spectrum Sensing," \emph{arXiv preprint arXiv:1905.01430}

\bibitem{Luo2020}
Z. Luo, S. Zhao, Z. Lu, Y. E. Sagduyu, and J. Xu, ``Adversarial Machine Learning Based Partial-model Attack in IoT," \emph{ACM  Workshop on Wireless Security and Machine Learning (WiseML)}, 2020.

\bibitem{Luo2021}
Z. Luo, Z. Shangqing, R. Duan, Z. Lu, Y. E. Sagduyu, and J. Xu, ``Low-cost Influence-Limiting Defense against Adversarial Machine Learning Attacks in Cooperative Spectrum Sensing," \emph{ACM  Workshop on Wireless Security and Machine Learning (WiseML)}, 2021.

\bibitem{Davaslioglu19}
K.~Davaslioglu and Y.~Sagduyu, ``Trojan Attacks on Wireless Signal Classification with Adversarial Machine Learning," \emph{IEEE DySPAN Workshop on Data-Driven Dynamic Spectrum Sharing}, 2019.

\bibitem{Shi2019}
Y. Shi, K. Davaslioglu, and Y. E. Sagduyu, ``Generative Adversarial Network for Wireless Signal Spoofing," \emph{ACM Conference on Security and Privacy in Wireless and Mobile Networks (WiSec) Workshop on Wireless Security and Machine Learning (WiseML)}, 2019.

\bibitem{Shi2021}
Y. Shi, K. Davaslioglu, and Y. E. Sagduyu, ``Generative Adversarial Network in the Air: Deep Adversarial Learning for Wireless Signal Spoofing," \emph{IEEE Transactions on Cognitive Communications and Networking}, Mar. 2021.

\bibitem{5Gbc}
Y. E. Sagduyu, T. Erpek, and Y. Shi, ``Adversarial Machine Learning for 5G Communications Security," \emph{arXiv preprint} arXiv:2101.02656, 2021.

\bibitem{Gunduz1}
M. Z. Hameed, A. Gyorgy, and D. Gunduz, ``Communication without Interception: Defense against Modulation Detection," \emph{IEEE Global Conference on Signal and Information Processing (GlobalSIP)}, 2019.

\bibitem{Gunduz2}
M. Z. Hameed, A. Gyorgy, and D. Gunduz, ``The Best Defense is a Good Offense: Adversarial Attacks to Avoid Modulation Detection," \emph{IEEE Transactions on Information Forensics and Security}, Sept. 2020.

\bibitem{Kim3}
B. Kim, Y. E. Sagduyu, K. Davaslioglu, T. Erpek, and S. Ulukus, ``How to Make 5G Communications `Invisible': Adversarial Machine Learning for Wireless Privacy," \emph{Asilomar Conference on Signals, Systems, and Computers}, 2020.

\bibitem{conventional1}
W. Xu, W. Trappe, Y. Zhang, and T. Wood, ``The Feasibility of Launching and Detecting Jamming Attacks in Wireless Networks," \emph{ACM International Symposium on Mobile Ad Hoc Networking and Computing (MobiHoc)}, 2005.

\bibitem{conventional2}
Y. E. Sagduyu, R. Berry, and A. Ephremides, ``Jamming Games in Wireless Networks with Incomplete Information," \emph{IEEE Communications Magazine}, Aug. 2011.


\bibitem{ModelInv}
M. Fredrikson, S. Jha, and T. Ristenpart, ``Model Inversion Attacks that Exploit Confidence Information and Basic Countermeasures," \emph{ACM SIGSAC Conference on Computer and Communications Security} 2015.

\bibitem{Shokri17}
R. Shokri, M. Stronati, C Song, and V Shmatikov, ``Membership Inference Attacks against Machine Learning Models," \emph{IEEE Symposium on Security and Privacy}, 2017.

\bibitem{Nasr18}
M.~Nasr, R.~Shokri, and A.~Houmansadr, ``Machine Learning with Membership Privacy using Adversarial Regularization," \emph{ACM Conference on Computer and Communications Security (CCS)}, 2018.

\bibitem{Yeom18}
S. Yeom, I. Giacomelli, M. Fredrikson and S. Jha, ``Privacy Risk in Machine Learning: Analyzing the Connection to Overfitting," \emph{IEEE Computer Security Foundations Symposium (CSF)}, 2018.

\bibitem{Salem18}
A. Salem, Y. Zhang, M. Humbert, P. Berrang, M. Fritz, and M. Backes, ``ML-Leaks: Model and Data Independent Membership Inference Attacks and Defenses on Machine Learning Models," \emph{Network and Distributed System Security Symposium (NDSS)}, 2018.

\bibitem{Leino19}
K.~Leino and M.~Fredrikson, ``Stolen Memories: Leveraging Model Memorization for Calibrated White-Box Membership Inference," \emph{arXiv preprint, http://arxiv.org/abs/1906.11798}, 2019.

\bibitem{Nasr19}
M. Nasr, R. Shokri, and A. Houmansadr, ``Comprehensive Privacy Analysis of Deep Learning: Passive and Active White-box Inference Attacks against Centralized and Federated Learning," \emph{IEEE Symposium on Security and Privacy}, 2019.

\bibitem{Song19}
L. Song, R. Shokri, and P. Mittal, ``Privacy Risks of Securing Machine Learning Models against Adversarial Examples,"
\emph{ACM Conference on Computer and Communications Security (CCS)}, 2019.

\bibitem{Song19-2}
L. Song, R. Shokri, and P. Mittal, ``Membership Inference Attacks Against Adversarially Robust Deep Learning Models," \emph{IEEE Security and Privacy Workshops (SPW)}, 2019.

\bibitem{Truex19}
S. Truex, L. Liu, M. E. Gursoy, L. Yu and W. Wei, ``Demystifying Membership Inference Attacks in Machine Learning as a Service," \emph{IEEE Transactions on Services Computing}, 2019.



\bibitem{Jia20}
J. Jia, A. Salem, M. Backes, Y. Zhang, and N. Z. Gong, ``MemGuard: Defending against Black-Box Membership Inference Attacks via Adversarial Examples," \emph{ACM SIGSAC Conference on Computer and Communications Security}, 2019.

\bibitem{Song20}
L. Song and P. Mittal, ``Systematic Evaluation of Privacy Risks of Machine Learning Models," \emph{USENIX Security Symposium}, 2021.

\bibitem{Gong21}
B. Hui, Y. Yang, H. Yuan, P. Burlina, N. Z. Gong, and Y. Cao, ``Practical Blind Membership Inference Attack via Differential Comparisons," \emph{Network and Distributed System Security Symposium (NDSS)}, 2021.

\bibitem{Dingfan}
D. Chen, N. Yu, Y. Zhang, and M. Fritz, ``GAN-Leaks: A Taxonomy of Membership Inference Attacks against Generative Models," \emph{ACM SIGSAC Conference on Computer and Communications Security (CCS)}, 2020.

\bibitem{PHY}
N. Wang, T. Jiang, S. Lv, and L. Xiao, ``Physical-Layer Authentication based on Extreme Learning Machine," \emph{IEEE Communications Letters}, July 2017.

\bibitem{Dyspan}
Y. Shi, K. Davaslioglu, Y. E. Sagduyu, W. C. Headley, M. Fowler, and G. Green, ``Deep Learning for Signal Classification in Unknown and Dynamic Spectrum Environments," \emph{IEEE International Symposium on Dynamic Spectrum Access Networks (DySPAN)}, 2019.

\bibitem{Monson}
X. Qiu, J. Dai, and M. Hayes, ``A Learning Approach for Physical Layer Authentication using Adaptive Neural Network," \emph{IEEE Access}, 2020.

\bibitem{Deepwifi}
K. Davaslioglu, S. Soltani, T. Erpek, and Y. E. Sagduyu, ``DeepWiFi: Cognitive WiFi with Deep Learning," \emph{IEEE Transactions on Mobile Computing}, Feb. 2021.



\bibitem{Shi17HST}
Y.~Shi, Y.E.~Sagduyu, and A.~Grushin, ``How to Steal a Machine Learning Classifier with Deep Learning," \emph{IEEE Symposium on Technologies for Homeland Security (HST)}, 2017.




\end{thebibliography}
\end{document}